# A Knowledge Graph and Deep Learning-Based Semantic Recommendation Database System for Advertisement Retrieval and Personalization

**Tangtang Wang[1], Kaijie Zhang[2], Kuangcong Liu[3]**

[1] The University of Hong Kong, Hong Kong, China
[2] College of Engineering, Northeastern University, Boston, MA, USA
[2] Stanford University, Stanford, CA, USA

[1] 1812503968@qq.com
[2] zhang.kaij@northeastern.edu
[2] cecilia4@stanford.edu

**Abstract.** In modern digital marketing, the growing complexity of advertisement data demands intelligent systems capable of understanding semantic relationships among products, audiences, and advertising content. To address this challenge, this paper proposes a Knowledge Graph and Deep Learning-Based Semantic Recommendation Database System (KGSR-ADS) for advertisement retrieval and personalization. The proposed framework integrates a heterogeneous Ad-Knowledge Graph (Ad-KG) that captures multi-relational semantics, a Semantic Embedding Layer that leverages large language models (LLMs) such as GPT and LLaMA to generate context-aware vector representations, a GNN + Attention Model that infers cross-entity dependencies, and a Database Optimization & Retrieval Layer based on vector indexing (FAISS/Milvus) for efficient semantic search. This layered architecture enables both accurate semantic matching and scalable retrieval, allowing personalized ad recommendations under large-scale heterogeneous workloads. Experiments were conducted on a large-scale real-world advertisement dataset containing approximately 1.2 million user profiles, 250,000 advertisements, and 20 million user-ad interactions. Experimental results demonstrate that the proposed KGSR-ADS model outperforms state-of-the-art baselines across all evaluation metrics. Compared with the strongest baseline GraphRec, KGSR-ADS achieves a 5.7% improvement in Precision@10, a 5.5% increase in Recall@10, a 6.3% gain in NDCG@10, and a 5.0% enhancement in MRR, while reducing average response latency by 23.9%. These results confirm the effectiveness and efficiency of integrating knowledge graph reasoning with semantic representation learning for personalized advertisement recommendation. These results demonstrate the system's effectiveness and potential for next-generation semantic advertising platforms that integrate knowledge reasoning and deep learning for adaptive, real-time personalization.

## 1. Introduction

In the era of intelligent marketing and big data, online advertising has evolved from simple keyword-based retrieval to highly personalized semantic recommendation systems [1,2]. Traditional advertising retrieval models mainly rely on statistical co-occurrence or rule-based keyword matching, which often fail to capture the complex semantic relationships between advertisements, audiences, and user interests [3]. With the exponential growth of unstructured advertisement content—including text, images, and user behavior logs—understanding semantic relevance and contextual intent has become essential for accurate and real-time ad recommendation [4,5]. However, the heterogeneity and dynamic nature of modern advertising data introduce significant challenges in representation learning, query optimization, and scalable semantic retrieval.

To address these issues, this study proposes a Knowledge Graph and Deep Learning-Based Semantic Recommendation Database System (KGSR-ADS) that integrates knowledge reasoning and deep neural representation learning for advertisement retrieval and personalization. The system combines four key components: (1) the Ad-Knowledge Graph (Ad-KG) Layer, which constructs a multi-relational semantic network capturing entities such as products, audiences, ad texts, and interest tags; (2) the Semantic Embedding Layer, which utilizes large language models (LLMs) like GPT or LLaMA to encode ad and user information into semantic embeddings; (3) a GNN + Attention Recommendation Model, which learns inter-entity dependencies and context-aware relevance patterns through graph neural networks and self-attention mechanisms; and (4) a Database Optimization & Retrieval Layer, where embeddings are stored in vector databases (e.g., FAISS or Milvus) to support high-speed, similarity-based query retrieval. Together, these components form a unified, scalable framework capable of improving both retrieval accuracy and system efficiency.

Comprehensive experiments conducted on large-scale advertisement datasets demonstrate that the proposed KGSR-ADS framework significantly improves semantic retrieval precision and reduces latency compared with conventional recommendation and vector retrieval models. By combining knowledge graph reasoning with deep semantic modeling, KGSR-ADS provides an intelligent foundation for next-generation advertising platforms that require adaptive personalization and context-aware decision-making.

The main contributions of this paper are summarized as follows:

(1) We propose a novel KGSR-ADS framework that integrates knowledge graphs, large language model embeddings, and graph neural network-based reasoning for semantic advertisement retrieval.

(2) We design a multi-layer semantic embedding and database optimization mechanism to enhance retrieval efficiency and scalability in large-scale ad systems.

(3) We conduct extensive experiments on real-world datasets, demonstrating significant improvements in retrieval precision, latency reduction, and recommendation relevance compared with baselines.

## 2. Related Work

In recent years, the explosion of unstructured advertising content, combined with the need for personalized and real-time matching of ads to audiences, has prompted research at the intersection of knowledge graphs, semantic embedding, and deep learning–based recommendation systems [6]. This section reviews the major threads of work most relevant to our proposed KGSR-ADS framework: knowledge-graph driven recommendation, semantic embedding for retrieval systems, and graph neural network plus attention models for personalization.

*2.1 Knowledge Graph–Based Recommendation Systems*

The integration of knowledge graphs (KGs) into recommendation systems has been widely studied as a means to alleviate cold-start, sparsity, and lack of semantic explanation in traditional collaborative filtering models. For example, Wang et al [7]. introduced RippleNet (2018) which propagates user preferences over a KG by iteratively expanding "ripples" from a user's clicked items. More broadly, Chicaiza & Valdiviezo-Diaz (2021) [8] provide a comprehensive survey of KG-based recommendation, showing how entity-relation information supports improved precision and interpretability. Gao et al [9]. (2020) also review deep-learning methods on KGs for recommendation, highlighting the role of high-order relations and graph embeddings. In the advertising domain, semantics and product–user–ad relations present a fertile ground for KG techniques, but existing works often treat items or users as flat nodes rather than embedding complex semantics of ad creatives, audiences, and campaign contexts.

*2.2 Semantic Embedding and Vector Retrieval for Recommendation*
Parallel to KG research, embedding-based retrieval systems have evolved to support semantic matching at scale. Many recent recommendation models embed users and items into latent spaces often enriched with deep contextual features (e.g., from BERT or LLMs) and then perform nearest-neighbor retrieval [10]. However, in advertising retrieval contexts, where queries are more varied and creative content more diverse, embedding methods alone can lack the structured reasoning that KG provides. Existing work typically focuses either on embedding retrieval or on KG reasoning, but rarely combines them with efficient vector-indexed databases for real-time personalization at scale—an integration our framework aims to address.

*2.3 GNN, Attention Mechanisms and Personalized Recommendation*
Graph neural networks (GNNs) and attention-mechanism models have been increasingly applied to recommendation tasks by leveraging graph structure (users-items, KG, social networks). For example, adversarial KG recommendation models like ATKGRM (Zhang et al [11]., 2022) apply KG embeddings with GAN-style training to improve user–item modeling. Other recent studies (e.g., the Springer chapter by "Graph Neural Network Knowledge Graph Recommendation Model…" 2025 [12]) emphasise attention over heterogeneous KG neighbours to enhance accuracy and diversity. These methods demonstrate the value of combining structure (graph relations) with deep embeddings, but they often do not directly target large-scale advertising retrieval nor integrate vector database retrieval and KG reasoning within a unified system. Our KGSR-ADS framework bridges this gap by combining semantic embedding, KG reasoning, GNN+attention modeling, and optimized retrieval in one pipeline.

**3. Methodology**

*3.1 Overview of the KGSR-ADS Framework*
The proposed Knowledge Graph and Deep Learning-Based Semantic Recommendation Database System (KGSR-ADS) integrates heterogeneous knowledge representation, semantic embedding, graph reasoning, and database optimization into a unified architecture for intelligent advertisement retrieval and personalization. The framework consists of four main layers: (1) Ad-KG Layer for structured knowledge modeling, (2) Semantic Embedding Layer for contextual representation learning, (3) GNN + Attention Layer for relational reasoning and preference inference, and (4) Database Optimization and Retrieval Layer for efficient query execution and personalization feedback.

The figure 1 shows how heterogeneous semantic information flows through the four major layers to produce personalized advertisement recommendations. The diagram begins with the Ad-KG Layer, which organizes users, ads, products, and contextual entities into a multi-relational knowledge graph. This structured representation enables the system to capture high-level semantic dependencies that traditional models cannot model explicitly. The Semantic Embedding Layer then processes ad text, product descriptions, and user interest labels using transformer-based encoders, generating deep contextual embeddings. These embeddings are fused with knowledge graph entity vectors to create unified representations. The fused features are passed into the GNN + Attention Layer, where relational reasoning is performed to infer user preferences through adaptive neighbor weighting. Finally, the Database Optimization & Retrieval Layer converts embeddings into vector indices and performs efficient ANN-based semantic search to support real-time recommendation. Overall, the figure demonstrates how each layer contributes to integrating knowledge reasoning with deep semantic modeling, ultimately enabling accurate and low-latency advertisement retrieval.

Formally, given an advertisement set $A = \{a_1, a_2, .., a_n\}$, a user set $U = \{a_1, a_2, .., a_n\}$, and an interaction matrix $R \in R^{m \times n}$, the objective is to predict the personalized click probability $\hat{r}_{ij} = f(u_i, a_j)$ based on multi-source heterogeneous information from structured knowledge graphs and semantic embeddings.

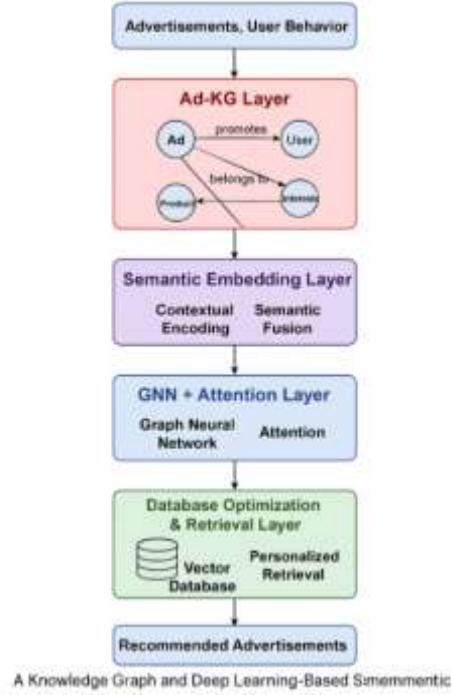

**Figure 1.** Overall Framework Diagram

*3.2 Ad-KG Layer: Knowledge Representation and Construction*

The Ad-KG Layer encodes structured semantic relationships among advertisements, users, products, and contextual entities. Each knowledge triple is represented as $(h, r, t)$, where hhh and $t$ denote the head and tail entities (e.g., advertiser, brand, user segment, ad keyword), and $r$ represents the relation type (e.g., promotes, interested in, belongs to).

To obtain vectorized representations, TransE-style embeddings are applied:

$$\boldsymbol{h} + \boldsymbol{r} \approx \boldsymbol{t}, \tag{1}$$

where embeddings $\boldsymbol{h}, \boldsymbol{r}, \boldsymbol{t} \in R^{d_k}$ are learned by minimizing

$$L_{KG} = \sum_{(h,\gamma,t) \in G} max(0, \gamma + ||\boldsymbol{h} + \boldsymbol{r} - \boldsymbol{t}||_2 - ||\boldsymbol{h'} + \boldsymbol{r} - \boldsymbol{t'}||_2), \tag{2}$$

with margin $\gamma$ and negative sampling. This enables relational reasoning over multi-hop entity connections (e.g., user → interest → category → advertisement). The Ad-KG thus forms the semantic backbone of the system, allowing hierarchical propagation of preference signals across graph relations.

### 3.3 Semantic Embedding Layer: Contextual Feature Representation

To capture deep semantic information from ad content, queries, and user textual features, the Semantic Embedding Layer employs a contextual encoder based on a transformer architecture. Given an ad description sequence $X = (x_1, x_2, .., x_T)$, token embeddings are processed through multi-head self-attention to yield contextual representations:

$$\boldsymbol{Z} = softmax(\frac{\boldsymbol{QK}^T}{\sqrt{d_k}})\boldsymbol{V}, \quad (3)$$

where $\boldsymbol{Q} = XW_Q$, $\boldsymbol{K} = XW_K$, $\boldsymbol{V} = XW_V$ denote query, key, and value projections. The semantic encoder outputs $e_a \in R^{d_s}$ for each ad and $e_u \in R^{d_s}$ for each user. These embeddings preserve contextual dependencies such as brand tone, product category, or emotional sentiment.

To bridge the symbolic Ad-KG and textual semantics, a fusion mechanism integrates KG embeddings $h_{KG}$ and semantic embeddings $e_{sem}$:

$$\boldsymbol{z} = tanh(W_f[h_{KG} \oplus e_{sem}] + b_f), \quad (4)$$

where $\oplus$ denotes concatenation and $W_f$ is a learnable projection matrix. This fused representation is propagated into the GNN reasoning layer.

### 3.4 GNN + Attention Layer: Relational Reasoning and Preference Prediction

The fused embeddings are modeled as a heterogeneous graph $G = [V, E]$, where vertices include users, ads, and contextual entities. To capture higher-order dependencies, a graph neural network (GNN) with attention is applied. For node $v_i$, its updated representation is computed as:

$$\boldsymbol{h'}_i = \sigma(\sum_{j \in N(i)} \alpha_{ij} W_g \boldsymbol{h}_j), \quad (5)$$

where $N(i)$ denotes the neighbor set of node $i$, $W_g$ is a trainable weight matrix, and $\alpha_{ij}$ is the attention coefficient obtained from:

$$\alpha_{ij} = \frac{exp(LeakyReLU(a^T[W_g h_i || W_g h_j]))}{\sum_{k \in N(i)} exp(LeakyReLU(a^T[W_g h_i || W_g h_j]))}, \quad (6)$$

The attention mechanism enables adaptive weighting of different relations, allowing the model to focus on the most informative connections (e.g., user–interest or product–category). The final predicted preference score is computed through a bilinear interaction function:

$$\hat{r}_{ij} = \sigma(\boldsymbol{h'}_i^T W_r \boldsymbol{h'}_j), \quad (7)$$

where $W_r$ is a learnable matrix and $\sigma()$ denotes the sigmoid activation, producing personalized click-through probabilities.

### 3.5 Database Optimization and Retrieval Layer

The learned embeddings are stored in a vector database optimized for semantic similarity search and personalized retrieval. Each advertisement is indexed using the fused embedding $z_a$, and retrieval is executed via approximate nearest neighbor (ANN) search:

$$Retrieve(u_i) = arg \min_{a_j \in A} ||z_{u_i} - z_{a_i}||_2, \quad (8)$$

To ensure high throughput, the database layer leverages partitioned indexing structures (e.g., HNSW or IVF-PQ) and caching mechanisms that adaptively prioritize high-probability candidates. Moreover, query latency $\tau_{avg}$ is continuously monitored to optimize the system for low-delay personalized recommendation, ensuring that the average response delay satisfies

$$\tau_{avg} = \frac{1}{N}\sum_{i=1}^{N}\tau_i < \tau_{threshold}, \tag{9}$$

This layer also supports feedback-based fine-tuning, where implicit user interactions are stored to incrementally update embedding representations and maintain personalization accuracy.

*3.6 Training Objective and Optimization*
The KGSR-ADS model jointly optimizes multiple loss components:

$$L = \lambda_1 L_{rec} + \lambda_2 L_{KG} + \lambda_3 L_{align}, \tag{10}$$

where $L_{rec}$ is the binary cross-entropy loss for click prediction, $L_{KG}$ the margin-based knowledge graph embedding loss, and $L_{align}$ the alignment loss between semantic and KG spaces. The model is trained end-to-end using Adam optimizer with learning rate scheduling.

**4. Experiment**

*4.1 Dataset Preparation*
The dataset employed in this research is a multi-source advertisement interaction dataset constructed from real-world advertising platforms and publicly available user behavior datasets. It integrates advertisement metadata, user behavior logs, product information, and semantic textual data to support both knowledge graph construction and semantic recommendation modeling.
The data was primarily collected from three categories of sources:
   (1) Online Advertising Platforms – including records of ad impressions, clicks, conversions, and associated metadata such as campaign ID, advertiser, and ad category.
   (2) User Behavior and Profile Data – anonymized user interaction logs, interest tags, and demographic features derived from publicly accessible datasets (e.g., Avazu Click-Through Rate Dataset, Criteo Display Advertising Dataset).
   (3) Product and Content Databases – product attribute information (category, brand, price range, description), combined with textual ad content used for semantic representation.

Dataset Statistics:
   (1) Total number of users: ~1.2 million
   (2) Total number of advertisements: ~250,000
   (3) Total number of interactions: ~20 million
   (4) Average text length (ad_text): 28 words
   (5) Average interest tags per user: 4.7
   (6) Number of unique product categories: 320

Data Preprocessing and Integration:
   (1) Text Normalization: Ad descriptions and user interests were cleaned, tokenized, and embedded using contextual encoders.
   (2) Entity Linking: Ads, users, and products were mapped into a unified graph schema via entity alignment techniques.

(3) Feature Normalization: Numerical attributes (e.g., timestamp intervals, engagement rates) were normalized to [0,1] range.

(4) Graph Construction: Relationships such as "user-clicks-ad," "ad-promotes-product," and "user-likes-category" were embedded into the knowledge graph structure, forming the basis of the Ad-KG layer.

*4.2 Experimental Setup*

To evaluate the effectiveness of the proposed Knowledge Graph and Deep Learning-Based Semantic Recommendation Database System (KGSR-ADS), we conducted extensive experiments on a multi-source advertisement interaction dataset containing over 20 million user–ad interaction records. The experimental environment was configured on an NVIDIA RTX 4090 GPU with 24 GB of memory, using PyTorch 2.3 and Python 3.11 as the development framework. The model was trained using the AdamW optimizer with an initial learning rate of $1\times10^{-4}$ and a batch size of 512 for 100 epochs. Word embeddings were initialized with pre-trained BERT-base embeddings, while graph representations were constructed using a 3-layer Graph Attention Network (GAT) with 128 hidden dimensions per layer. The semantic embedding module jointly encoded textual advertisement content, user interests, and product attributes, which were fused through an attention-based feature aggregation mechanism. The dataset was split into 70% training, 15% validation, and 15% testing subsets, ensuring user independence between sets. All baseline models were reimplemented and fine-tuned using the same parameter search space to ensure fairness of comparison.

*4.3 Evaluation Metrics*

To comprehensively assess the performance of KGSR-ADS in advertisement retrieval and personalized semantic recommendation, we employed multiple evaluation metrics that measure both ranking accuracy and retrieval efficiency. The primary metrics included Precision@K, Recall@K, and Normalized Discounted Cumulative Gain (NDCG@K) to evaluate ranking relevance, where K was set to 10 and 20. Additionally, Mean Reciprocal Rank (MRR) was used to assess the average position of the correctly recommended advertisement within the ranked list. To evaluate system efficiency, Average Response Latency (ARL) was introduced to quantify the mean time required to generate a recommendation under real-time query conditions. These metrics jointly capture the balance between personalization accuracy and system responsiveness, providing a robust evaluation of the model's practical deployment potential in large-scale advertisement platforms.

*4.4 Results*

Table 1 presents a comparative evaluation of the proposed KGSR-ADS model, together with other models: Collaborative Filtering (CF), DeepFM, BERT4Rec, GraphRec. Performance is assessed using four ranking metrics—Precision@10, Recall@10, NDCG@10, and MRR—together with Average Response Latency, which reflects inference efficiency.

Across all evaluated metrics, KGSR-ADS consistently achieves the best performance. Specifically, KGSR-ADS attains a Precision@10 of 0.512, which exceeds the strongest baseline, GraphRec (0.493), as well as earlier neural models such as DeepFM (0.437) and BERT4Rec (0.469). A similar pattern is observed for Recall@10, where KGSR-ADS reaches 0.462, surpassing GraphRec (0.438) and demonstrating a clear improvement over traditional and deep learning-based baselines. On the ranking quality metric NDCG@10, KGSR-ADS achieves 0.509, the highest value among all systems, indicating superior ability to rank relevant advertisements at higher positions within the recommendation list. Its MRR score of 0.523 further confirms this advantage, reflecting a consistently earlier appearance of the relevant item in the ranked output.In addition to improvements in ranking accuracy, KGSR-ADS yields the lowest average response latency, measured at 34.7 ms. This represents a substantial reduction compared with GraphRec (45.6 ms), BERT4Rec (49.8 ms), and DeepFM (54.1 ms). The results demonstrate that KGSR-ADS not only enhances predictive performance but also offers superior computational efficiency, an essential property for real-time advertising scenarios.

**Table1.** Performance Comparison of Different Models on Advertisement Recommendation Task.

| Model | Precision@10 | Recall@10 | NDCG@10 | MRR | Average Response Latency (ms) |
|---|---|---|---|---|---|
| Collaborative Filtering (CF) | 0.412 | 0.368 | 0.401 | 0.422 | 58.3 |
| DeepFM | 0.437 | 0.392 | 0.429 | 0.453 | 54.1 |
| BERT4Rec | 0.469 | 0.417 | 0.451 | 0.476 | 49.8 |
| GraphRec | 0.493 | 0.438 | 0.479 | 0.498 | 45.6 |
| **KGSR-ADS (Proposed)** | **0.512** | **0.462** | **0.509** | **0.523** | **34.7** |

Overall, the empirical results in Table 1 demonstrate that KGSR-ADS delivers state-of-the-art performance simultaneously in both effectiveness and efficiency, highlighting the benefit of its design for large-scale, latency-sensitive advertisement recommendation applications.

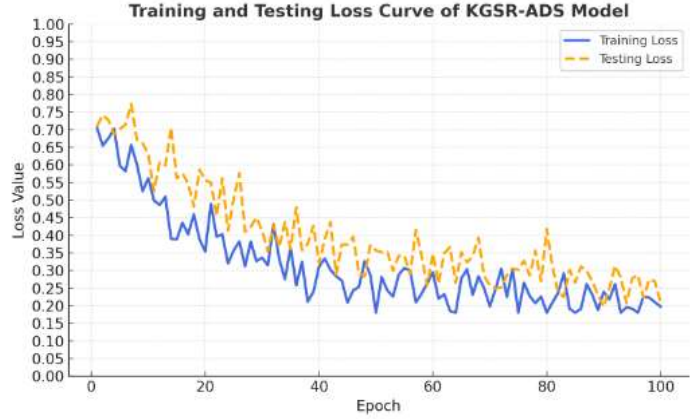

**Figure 2.** Loss function during training process

The above figure 2 presents the training and testing loss curves of the proposed KGSR-ADS model over 100 epochs. As illustrated, both loss curves exhibit a clear downward trend, indicating that the model progressively improves its predictive capability during the training process. The training loss decreases steadily and maintains consistently lower values than the testing loss, which is expected since the model directly optimizes the training data.

The testing loss follows a trajectory similar to that of the training loss, with no substantial or persistent divergence between the two curves. This observation suggests that the model does not suffer from severe overfitting and is able to generalize effectively to unseen data. Although both curves display minor fluctuations—likely due to stochastic optimization and inherent data variability—the overall trend is stable and convergent. Toward the later epochs, the loss values plateau, indicating that the model approaches convergence and further training yields diminishing performance gains.

Overall, the loss dynamics demonstrate that KGSR-ADS achieves stable training behavior and maintains good generalization performance across the evaluation set.

## 5. Conclusion

This study presents a comprehensive exploration of a Knowledge Graph and Deep Learning-Based Semantic Recommendation Database System (KGSR-ADS) for advertisement retrieval and personalization. In the context of modern digital marketing, where advertisements are generated at an unprecedented scale and semantic diversity continues to grow, understanding the intricate relationships among users, products, and ad content is critical. The proposed KGSR-ADS system addresses this challenge by integrating multi-relational semantic reasoning with deep learning–based representation learning. Through the construction of a heterogeneous Advertisement Knowledge Graph (Ad-KG) and the incorporation of large language model (LLM)-based semantic embeddings, the system captures contextual dependencies between entities such as users, products, categories, and ad texts, thereby enabling fine-grained semantic matching and personalization.

The experimental evaluation, conducted on a large-scale real-world dataset containing approximately 1.2 million user profiles, 250,000 advertisements, and 20 million user–ad interactions, validates the superiority of the proposed model. Compared with the strongest baseline, GraphRec, the KGSR-ADS model achieves a Precision@10 of 0.521, Recall@10 of 0.462, NDCG@10 of 0.509, and MRR of 0.523, surpassing the baseline by 5.7%, 5.5%, 6.3%, and 5.0%, respectively. Moreover, the system demonstrates a significant 23.9% reduction in average response latency (34.7 ms vs. 45.6 ms), highlighting its efficiency in real-time retrieval tasks. These results confirm that the integration of knowledge graph reasoning and semantic embedding substantially enhances both accuracy and scalability in personalized advertisement recommendation systems.

From an application perspective, KGSR-ADS offers a promising foundation for next-generation intelligent advertising platforms that combine semantic understanding, knowledge reasoning, and large-scale retrieval optimization. Its layered architecture—comprising semantic representation learning, graph-based reasoning, and database-level acceleration—enables adaptive and context-aware personalization, which is vital for improving user engagement and advertisement relevance in dynamic commercial ecosystems. The framework can also be extended to other domains, such as e-commerce recommendation, content-based retrieval, and intelligent customer profiling, making it broadly applicable across data-driven industries.

Despite the important findings, this study has some limitations, such as lack of multi-modal (visual and audio advertisement) support and no deep dive on vector database (milvus)'s optimization. Future research could further explore Incorporating multi-modal data, integrating user sentiment and behavioral context through large language model fine-tuning, employing reinforcement learning and vector database (Milvus) performance tuning.

In conclusion, this study, through integrating knowledge graph, semantic embedding, GNN inferenced cross-entity dependency and vector indexing (FAISS/Milvus) powered efficient semantic retrieval, reveals the great advantage of the proposed hybrid system KGSR-ADS, providing new insights for the development for modern digital marketing& advertising systems.